# CURVED EXTENDED SUPERSPACE FROM YANG-MILLS THEORY À LA STRINGS

W. Siegel[1]

*Institute for Theoretical Physics*
*State University of New York, Stony Brook, NY 11794-3840*

## ABSTRACT

The massless superfield content of four-dimensional compactifications of closed superstrings with extended (N=2, 3, or 4) supersymmetry is derived by multiplying two (N=0, 1, or 2) Yang-Mills multiplets. In some cases these superfields are known, and the low-energy actions are determined from the fact that the compensator (dilaton) supermultiplets occur quadratically classically. In the other cases these superfields suggest new formulations of extended superspace theories.

---

[1] Internet address: siegel@insti.physics.sunysb.edu.

# 1. INTRODUCTION

## 1.1. General principles

Low-energy actions of string theories can be derived directly from field theory by incorporating "T-duality" [1] as a spontaneously broken symmetry [2]. T-duality is the symmetry which, in string theory and its low-energy limit, mixes the metric tensor with the antisymmetric tensor and, in the heterotic string, the Yang-Mills vectors. (In string language, it rotates the left- and right-handed components of the string coordinate $x$ as a 26+26, 10+10, or 26+10 component vector. Although T-duality is normally described as a symmetry of solutions independent of some coordinates, in our description the maximal T-duality can be made manifest in the action, is broken by the choice of 26- or 10-dimensional hypersurface, and is partially restored for such solutions, similar to a high-energy limit. In this paper "T-duality" will always refer to this extended definition.) Such low-energy actions are unique up to the usual ambiguitites of compactification. This method has the advantage of giving the low-energy information of strings without using the full machinery needed for calculating the effects of the massive states. Another advantage is that the usual four-dimensional superfield methods can be applied, avoiding questions of which formulation (Ramond-Neveu-Schwarz, Green-Schwarz, Berkovits) should be used to describe the corresponding superstring variables. In particular, it is straightforward to determine the off-shell formulation of the low-energy field theory. In this paper we will apply some of the properties following from T-duality to find some of these off-shell superspace formulations. The two main principles we will use are well known general properties of string theory:

(1) *The first-quantized Hilbert space of* (the massless fields of) *the closed string is the direct product of the Hilbert spaces of* (the massless fields of) *two open strings.* This factorization holds off shell as well as on shell, so that it applies to auxiliary fields as well. In the manifestly T-dual formulation of field theories corresponding to the low-energy limit of superstrings, this follows from an appropriate gauge choice [2]. In string language, it follows from the assumption of the existence of a corresponding string field theory, but it also can be derived from the more general assumption of Becchi-Rouet-Stora-Tyutin first-quantization of strings, since it is a direct consequence of the separability of this BRST operator in terms of left- and right-handed degrees of freedom. Furthermore, the massless sector of any open string is just a vector or (for open superstrings) vector multiplet in the critical dimension, which reduces upon compactification to a vector (multiplet) plus matter defined by the compactification. Unlike supergravity multiplets, whose off-shell content can be ambiguous because of their reducibility, vector multiplets are unambiguous. Thus, the structure of the massless sector of the closed string follows from the known structure of the massless sectors of the two open strings, plus the properties of the compactification. This off-shell field content includes not only the fields themselves, but also their gauge transformations (ghosts) and gauge fixing (antighosts). This principle is thus enough to identify the superfield content of the theory, in terms of prepotentials. It also determines which off-shell version of supergravity is used by string theory [3].

(2) *The dilaton multiplet(s) appears homogeneously of order 2 in the field theory action* (and thus homogeneously of order $2 \times (1 - \text{loops})$ in the field theory effective action). The field to which we refer as the "dilaton" is the scalar density that is invariant under T-duality. In the manifestly T-dual formulation, this principle is a consequence of the facts that the dilaton multiplets are the only densities around with which to construct actions. Using known properties of the superspace formulation of these multiplets, this principle then provides restrictions on the form of the action. If matter from the Calabi-Yau sector is ignored, it is sufficient to determine the low-energy action uniquely. In the cases where these prepotentials are familiar (N=0,1, and type II N=2), we construct these actions explicitly. In the remaining cases (N=3,4, and heterotic N=2), this approach gives new off-shell formulations of extended supergravity. It should be possible to complete these new formulations by incorporating the field content of the prepotentials into the constraints on the curved-superspace covariant derivatives, or more directly by constructing the corresponding covariant derivatives of the manifestly T-dual formulation (as for N=0,1 in [2]).

## 1.2. Nature of dilaton

This dilaton coupling in string language also follows from T-duality: Since T-duality mixes all the physical vertex operators, and the dilaton is invariant, the dilaton vertex operator can be expressed in terms of just world-sheet ghosts, which generate a world-sheet curvature coupling upon being integrated out. The fact that they couple to ghosts means that the dilaton multiplets are the ones that follow from considering the direct product of the ghost sectors of the two open-string Hilbert spaces.

Such states are a general property of field theory: If any representation of the Poincaré group is defined by adding two commuting and two anticommuting dimensions to the light-cone, then the fields that appear in the minimal gauge-invariant action are those that are singlets under the Sp(2) symmetry that rotates the two anticommuting dimensions. In string language, this is a symmetry that rotates the world-sheet ghost(s) and antighost(s). Thus, the dilaton and related states are those that couple to the Sp(2) singlet combination of the direct product of open-string vector (multiplet) ghosts with open-string vector (multiplet) antighosts.

More explicitly, all fields, upon adding extra dimensions to the light-cone representations of SO(D−2), become representations of OSp(D−1,1|2). In particular, an OSp vector $A_i$ contains the physical D-vector $A_a$ and the two ghosts $A_\alpha$. The direct product of two such vectors is then a tensor $T_{ij}$. This decomposes as usual into a supertraceless graded symmetric tensor, the supertrace, and a graded antisymmetric tensor. These separate tensors include all the usual components and ghosts: The supertraceless symmetric tensor includes the metric tensor and its trace and their ghosts, the antisymmetric tensor includes that gauge field and its ghosts and ghosts for ghosts. This separation is the one for which the (gauge-fixed) kinetic terms are diagonal [3]. However, for purposes of T-duality, and for writing $\sigma$ models for strings, it is more useful not to make this separation, since T-duality affects only the SO(D−1,1) part of each OSp index, which couples to $\partial x$ in the string, while the Sp(2) part of each index couples to the ghosts $b, c$. Thus, in this OSp language it is clear which fields are physical, etc.: The OSp supertrace $T^i{}_i$ is the physical scalar, which appears when the action is written in diagonal form, while the Sp(2) trace $T^\alpha{}_\alpha$ is the T-duality-invariant dilaton, which appears, with wrong-sign kinetic term because of its ghost coupling, when the action is written in string $\sigma$-model form.

We can then interpret the scalar density we are calling "dilaton" as the "compensating" scalar for dilatations. (This is also the original definition of "dilaton" from before the days of string theory.) This unphysical conformal mode of ordinary gravity always appears with negative metric in the action. Thus, the position of the physical and compensating fields has been reversed from those normally used in field theory [3]. This switch can be performed (or undone) by a field redefinition; however, a more useful method, especially in supergravity, is to begin with (super)gravity as conformal (super)gravity coupled to a scalar (multiplet) with wrong-sign kinetic term (the compensator), and couple to further, physical fields. (In this context "scalar multiplet" refers to any supersymmetric multiplet containing a scalar.) In this approach the action is invariant under local Weyl (super)scale transformations. The normal gauge for this invariance then results from gauging away the compensator (multiplet), while the "string gauge" comes from gauging away some physical field(s).

Two features of this dilaton have caused confusion in the literature: (1) In a Weyl scale gauge, the form of the torsions and curvatures of the superspace covariant derivatives depends on the type of scalar multiplet that has been gauged away. This is because this multiplet has been eaten by the conformal supergravity multiplet, so its fields show up there, as torsions and curvatures. On the other hand, the type of off-shell supergravity being used is defined by the type of scalar multiplet used as the compensator. Although these two properties are related in the normal Weyl gauge where it is the compensator that is gauged away, they are not generally related in the string gauge, where a physical multiplet is gauged away. In the $\sigma$-model approach, this means the type of off-shell supergravity can be determined only by considering the coupling to the world-sheet ghosts or curvature. (2) The "dilaton" that counts loops is the compensator, not the physical scalar.

## 1.3. Outline

The general picture is then to consider the physical and ghost fields of two vector multiplets, take their direct product, and identify the physical and ghost fields of the corresponding closed string. Specifically, physical $\otimes$ physical gives physical, physical $\otimes$ ghost and ghost $\otimes$ physical give ghost, and ghost $\otimes$ antighost − antighost $\otimes$ ghost gives compensating. (The other ghost $\otimes$ ghost combinations give ghosts for ghosts.) One way to understand this is in terms of first-quantized BRST: The closed-string BRST operator can be divided into left-handed and right-handed parts; this analysis of the Hilbert space then follows from considering states of different ghost number. The ghost states are also useful because they give the (linearized) gauge transformations and gauge conditions. Generally, the fields resulting from these products are reducible representations: In particular, we will examine the (super)spin content of the physical fields to break it into conformal (super)gravity plus matter. We will denote the closed string resulting from the direct product of an N=m open string with an N=n open string as N=(m,n), with a total of m+n supersymmetries, where m,n=0,1,2. (A similar notation was used in [4]. N=3 vector multiplets can also be described in harmonic superspace [5], but it is not clear what multiplets the ghosts represent, since there is no multiplet of any kind with only spins <1 for N>2, so we will not consider m or n=3 further here.)

In the following section we will review this method for the bosonic string. In this case the analysis should be fairly familiar from BRST quantization of this string [6], but the method will be described in a way which is not dependent on string theory, except for the two assumptions stated above. (In the bosonic case, T-duality is also needed to fix a relative coefficient.) This will allow a straightforward generalization to superstrings, while avoiding questions of how such superstrings are formulated, whether they are covariantly quantizable, whether supersymmetry is manifest, etc.

In section 3 we review the application of this approach to the 4D N=1 compactification of the heterotic superstring. This has also appeared previously: The identification of the superfield content of this "16−16" supergravity appeared in [7], while the complete nonlinear action was given in [8]. We applied this method to rederive the linearized theory and show the theory described a physical tensor multiplet coupled to old minimal supergravity in [3]. (Although it has sometimes been claimed to be instead a chiral scalar multiplet coupled to new minimal supergravity, this is clearly incorrect, since then the antisymmetric tensor would have to be auxiliary, whereas it is actually the usual physical field that comes from vector $\otimes$ vector, as in the bosonic case.) The manifestly N=1 supersymmetric 4D string that directly leads to this description was discovered by Berkovits [9], and applications to sigma-model calculations are being investigated [10]. We also give a derivation of an explicit expression for the Chern-Simons contribution to the N=1 tensor multiplet field strength as a Chern-Simons superform in terms of the potentials. Its existence is guaranteed by the background-field method and the existence of higher-loop divergences. (A less geometric form, as a parametric integral of the field equations in terms of explicit prepotentials, appeared in [8].)

The next section discusses low-energy limits of N=2 strings. For the type II N=2 string, this approach directly leads to the superfield formulation originally proposed by Rivelles and Taylor [11], and effectively extended to the nonlinear level by tensor calculus [12]. These results also follow from the type II Berkovits string [13], with left- and right-handed N=2 world-sheet supersymmetries. (Applications of these results to describe the general form of the effective action, using N=2 superspace methods, are also given in [13].) The heterotic N=2 string implies a new formulation of N=2 supergravity, which we describe only at the linearized level.

In the final section we consider N>2 strings, also linearized. The asymmetric type II N=3 string allows not only a comparison with the N=3 harmonic superspace formulation of super Yang-Mills, but also suggests such a formulation for supergravity. The results for N=4 type II strings suggest a new formulation of N=4 harmonic superspace.

## 2. N=0
## 2.1. Open ⊗ open and BRST

The simplest example is the bosonic string in the critical dimension. The massless states of the open string are described by a vector, whose ghost and antighost are scalars. The physical sector is then given by vector ⊗ vector, which decomposes into traceless symmetric metric tensor (conformal graviton), antisymmetric tensor ("axion"), and the physical scalar, which appears as the trace of the metric tensor, but which we have separated out:

$$A_{La} \otimes A_{Rb} = h_{ab} + b_{ab} + \eta_{ab}\chi$$

(To be more precise, we should really write this equation for a basis of first-quantized states.) The off-shell content is described by spins $1 \otimes 1 = 2 \oplus 1 \oplus 0$, while the on-shell content is described by helicities $(\pm 1) + (\pm 1) = (\pm 2) \oplus 0 \oplus 0$.

Similarly, for the ghosts (gauge parameters):

$$C_L \otimes A_{Ra} = \epsilon_a + \zeta_a, \qquad A_{La} \otimes C_R = \epsilon_a - \zeta_a$$

The explicit linearized BRST (gauge) transformations are then given by:

$$QA_a = \partial_a \lambda, \qquad Q(A_{La} \otimes A_{Rb}) = (QA_{La}) \otimes A_{Rb} + A_{La} \otimes QA_{Rb}$$
$$\Rightarrow \qquad Qh_{ab} = \partial_{(a}\epsilon_{b)} - \tfrac{2}{D}\eta_{ab}\partial_c\epsilon^c, \qquad Qb_{ab} = \partial_{[a}\zeta_{b]}, \qquad Q\chi = \tfrac{2}{D}\partial_a\epsilon^a$$

(Since we are actually constructing the fields in terms of first-quantized states, these transformations are given by the first-quantized BRST operator, but are equivalent to the usual second-quantized expressions, linearized.)

Finally, the dilaton is

$$C_L \otimes \tilde{C}_R - \tilde{C}_L \otimes C_R = \phi$$

In general, the (linearized) transformation law of the antighost is the "dual" of the transformation law of the gauge field. (The two transformations are generated by the same term in the first-quantized BRST operator.) In this case:

$$QC = 0, \qquad Q\tilde{C} = \partial_a A^a$$

(Without loss of generality, we use the form of the BRST transformation resulting from eliminating the Nakanishi-Lautrup auxiliary field by its field equation in the Fermi-Feynman gauge.) The transformation of the dilaton then follows:

$$Q\phi = \partial_a \epsilon^a$$

The dilaton is thus a density under general coordinate transformations. This is true in string theory if the dilaton is introduced through coupling to worldsheet ghosts, and follows from classical considerations, as does its invariance under T-duality. However, if the ghosts are integrated out at one (worldsheet) loop to give coupling to the worldsheet curvature, then the density and T-duality properties of the dilaton are also one-loop effects, and depend on the definition of the measure. Thus, ghost coupling simplifies the dilaton's properties by making them classical, and hence measure-independent. (In the superstring case, the measure ambiguity is also resolved by supersymmetry, since superfields satisfying ordinary chirality conditions and with nontrivial R-symmetry weight are necessarily densities.)

## 2.2. Nonlinearity

By generalizing the gauge transformations to fully nonlinear gravity in the standard way, we find the usual version of the low-energy effective action, but written in terms of the T-duality-invariant dilaton [2], allowing for a cosmological constant $\lambda$ (as would appear in subcritical dimensions):

$$S = -\int d^D x \, \phi (4\Box + R + \tfrac{1}{12} H^{abc} H_{abc} + \lambda)\phi.$$

(The relative coefficients of these terms can be determined by T-duality. In the supersymmetric cases, they are already fixed by supersymmetry.) The string coupling appears as the inverse of the vacuum value of the dilaton $\phi$. If all fields except the vierbein $e_a{}^m$ are represented with flat indices, then the square root of the string tension appears as the vacuum value of the vierbein, since then a spacetime derivative can appear only in the combination $e_a{}^m \partial_m$. Here $\phi$ is related to the more common form $\varphi$ of the dilaton field used in string theory by

$$\phi = (-g)^{1/4} e^{-\varphi}$$

The fact that this is the T-duality-invariant combination follows from the fact that $\phi$ must soak up the $\sqrt{-g}$ measure, since $\sqrt{-g}$ is not T-duality invariant.

At this point we have not yet made the $h_{ab} + \eta_{ab}\chi$ separation; this we now perform with full nonlinearity by the Weyl rescaling

$$g^{mn} \to \chi^2 g^{mn}$$

(We do not scale $\phi$, and leave $\phi$-dependence out of the metric rescaling, so that $\phi$ stays out of T-duality transformations. Also, $b_{mn}$ is unscaled to preserve gauge transformations.) The result is

$$S = -\int d^D x \left\{ \phi^2 \chi^2 R + (D-1)(\phi\nabla\chi)^2 - \left[\phi^{-1}\nabla(\phi^2\chi)\right]^2 + \tfrac{1}{12}\phi^2\chi^6 H^{abc}H_{abc} + \lambda\phi^2 \right\}.$$

In general relativity one normally breaks the scale invariance introduced by this rescaling by gauge-fixing $\phi = (-g)^{1/4}$ ($\varphi = 0$):

$$S = -\int d^D x \sqrt{-g}\left\{ \chi[-(D-2)\Box + R]\chi + \tfrac{1}{12}\chi^6 H^2 + \lambda \right\}$$

The major difference from the original action is the change in sign for the scalar kinetic term: The fact that $\chi$ now appears with the right sign identifies it as the physical scalar. To eliminate all scalar dependence for the Einstein term we can instead choose the slightly modified gauge $\phi = (-g)^{1/4}\chi^{-1}$:

$$S = -\int d^D x \sqrt{-g}[R + \tfrac{1}{12}\chi^4 H^2 + (D-2)(\nabla \ln \chi)^2 + \lambda\chi^{-2}].$$

Of course, one can also return to the form of the action before Weyl rescaling by instead choosing the string gauge $\chi = 1$.

## 3. N=1

### 3.1. Versions of off-shell supergravity

Most of the ambiguity in auxiliary fields in supersymmetric theories in four dimensions is related to the choice of representing helicity zero in terms of either scalar or second-rank antisymmetric tensor fields. In supergravity, this ambiguity is the choice of off-shell representation of compensators. For N=1 and 2 supergravity (and probably also 3 and 4), the only compensator and matter multiplets of interest are those that can be represented as differential forms in superspace, which include as potentials the 0-form (scalar multiplet), 1-form (vector multiplet), and 2-form (tensor multiplet), as well as multiplets that are completely auxiliary (3-form and 4-form). These multiplets are related by gauge invariance: As in the nonsupersymmetric case, the $n$-form gauge field (potential) has an $n-1$-form gauge parameter (and ghost) and $n+1$-form field strength, and both these relationships are expressed simply in terms of the exterior derivative.

They are also related by on-shell duality, which switches Bianchi identities with field equations: In D dimensions, an $n$-form field strength is Hodge-dual to a $(D-n)$-form field strength, so an $n$-form gauge field is dual to a $(D-n-2)$-form gauge field. Finally, they are related by off-shell duality, which switches Bianchi identities $dF=0$ with gauge conditions $d^*A = 0$ (where "*" is the Hodge dual): Thus an $n$-form gauge field is dual to a $(D-n-1)$-form gauge field. For example, in the bosonic case, in four dimensions, a 0-form (scalar) is on-shell dual to a 2-form gauge field, which also describes helicity 0 on shell. However, off shell the 2-form describes spin 1, and is off-shell dual to the 1-form (vector) gauge field. Similar remarks apply to the prepotentials and "reduced" field strengths in the supersymmetric case, although the relations are then not simply in terms of exterior derivatives.

Explicitly, we have potentials written as superforms $A = dz^M \wedge dz^N \wedge ...A_{...NM}$, with gauge transformations $\delta A = d\lambda$, field strengths $F = dA$, and Bianchi identities $dF = 0$. After applying appropriate constraints on the $F$'s, and solving them on the $A$'s, the superforms are reduced, and a new "$d$" is defined as the operation that appears in the gauge transformations, expressions for the field strengths, and Bianchi identities for these reduced forms. For example, for 4D N=1 we have [14]:

| rank | reduced $A$ | reduced $dA$ | superhelicity | superspin |
|---|---|---|---|---|
| 0 | $\phi$ | $i(\bar\phi - \phi)$ | $\frac{1}{4}$ | 0 |
| 1 | $V$ | $i\bar{d}^2 d_\alpha V$ | $\frac{3}{4}$ | $\frac{1}{2}$ |
| 2 | $\phi_\alpha$ | $\frac{1}{2}(d_\alpha \phi^\alpha + \bar{d}_{\dot\alpha} \bar\phi^{\dot\alpha})$ | $\frac{1}{4}$ | $\frac{1}{2}$ |
| 3 | $V$ | $\bar{d}^2 V$ | — | 0 |
| 4 | $\phi$ | 0 | — | — |

where $\phi$ and $\phi_\alpha$ are chiral, while $V$ is real. (Here we use generic symbols for these reduced forms, since they may be gauge fields, field strengths, or Bianchi identities. For example, $\phi_\alpha$ may represent the gauge field for the tensor multiplet, the field strength $W_\alpha$ for the vector multiplet, or the Bianchi identitiy for the scalar multiplet.) The superhelicity is the average helicity of the physical states in the multiplet (with complex conjugate states of the opposite sign superhelicity), while the spins of the off-shell degrees of freedom (physical and auxiliary, but not gauge) are given by the superspin $\otimes$ the spins of the superspin-0 multiplet. In general, a superspin-0 multiplet has a chiral scalar field strength (which also satisfies a type of "reality" condition for even N), so it has spins as high as N/2.

From this table we can read the expressions for the (linearized) fields strengths $W_\alpha$ of the vector multiplet $V$, and $G$ of the tensor multiplet $\phi_\alpha$, as well as the corresponding gauge transformations and Bianchi identities. (The field strength $G$ of the tensor multiplet is a real linear superfield containing the vector field strength of a second-rank antisymmetric tensor gauge field [15], and should not be confused with the complex linear superfield, which contains no gauge fields [16].) Furthermore, the field equations follow from duality (where $\phi_\alpha \to i\phi_\alpha$ for rank 2). The actions that give these field equations are just the integrals of the squares of the field strengths, over $d^2\theta$ if chiral or $d^4\theta$ if real. Although the tensor and vector multiplets both have superspin $\frac{1}{2}$ (they are off-shell dual), the bosonic components of even engineering dimension in one have odd dimension in the other, so the two multiplets are easy to distinguish. They also differ in their superhelicities on shell. (Superspin 1 describes the multiplet with maximum spin $\frac{3}{2}$, which can be included only when describing N>1 supersymmetry.)

The only relevant multiplet not included among the forms is supergravity itself. Conformal supergravity is superspin $\frac{3}{2}$ off shell, described by a vector superfield $U^a$. Supergravity is described by starting with the conformal action for a multiplet containing a scalar (e.g., the scalar multiplet $\phi$ or tensor multiplet $\phi_\alpha$), and covariantizing with respect to conformal supergravity, as the generalization of the bosonic case. On shell, the combination of these two multiplets describes superhelicity $\frac{7}{4}$. The actions for pure old-minimal and new-minimal supergravity are the covariantization with respect to conformal supergravity of the wrong-sign conformal actions

$$S_{OM} = \int d^4x d^4\theta \, \bar\phi\phi + \left( \int d^4x d^2\theta \, \lambda\phi^3 + h.c. \right), \qquad S_{NM} = -\int d^4x d^4\theta \, G \ln G$$

Unlike the nonsupersymmetric case, the action of supergravity coupled to matter cannot always be written as a pure supergravity action plus matter terms after some field redefinition. The action is the covariantization with respect to conformal supergravity of some conformal matter action, with the kinetic term of one (superspin-0) multiplet having the wrong sign. Like the supersymmetric case, local scale invariance can be used to gauge away one multiplet; that multiplet is then effectively absorbed into the supergravity prepotential $U^a$. For example, we can choose a gauge $G = 1$ for some tensor multipet $G$. Conformal supergravity also has a local U(1) (R-)symmetry (U(N) for N-extended supergravity, at least for N≤4); combined with scale invariance, it can be used to instead gauge $\phi = 1$ for some scalar multiplet $\phi$. As for the bosonic case, there is freedom in which multiplet to gauge away; unlike that case, the field content of the supergravity sector is dependent on what type of multiplet has been scaled away (scalar or tensor), and not on what type of multiplet the compensator is.

## 3.2. Low-energy superstrings

The open $\otimes$ open analysis applied to the bosonic string generalizes straightforwardly to the 4D N=1 heterotic string. Our starting point is again the vector multiplets: For the bosonic open string, the vector $A_a$ just described is now accompanied by scalars $\varphi^I$ resulting from the compactification from 26 to 10 dimensions. On the other hand(edness), we have the 4D N=1 vector multiplet, described by a real superfield $V$ with a chiral ghost $\Lambda$ (and its antichiral complex conjugate $\bar{\Lambda}$) and chiral antighost $\tilde{\Lambda}$ (and $\bar{\tilde{\Lambda}}$):

$$QV = \Lambda + \bar{\Lambda}, \qquad Q\Lambda = 0, \qquad Q\tilde{\Lambda} = \bar{d}^2 d^2 V$$

The physical closed string fields are then a real vector N=1 superfield and real scalar superfields:

$$V \otimes A_a = U_a, \qquad V \otimes \varphi^I = V^I$$

These prepotentials are in the string gauge (for superscale transformations). This means the supergravity prepotential at this point describes a reducible multiplet, consisting of conformal supergravity plus the multiplet that was gauged away. The decomposition of the reducible multiplet described by $U_a$ is more subtle than in the bosonic case, but the basic idea is simple: $V$ describes a vector multiplet with superspin $\frac{1}{2}$, while $A_a$ describes a vector of spin 1; taking their direct product gives superspins $\frac{1}{2} \otimes 1 = \frac{3}{2} \oplus \frac{1}{2}$, where superspin $\frac{3}{2}$ describes the irreducible N=1 conformal supergravity multiplet, while this superspin $\frac{1}{2}$ describes a tensor multiplet. A similar analysis of superhelicities gives the analogous result $(\pm\frac{3}{4}) + (\pm 1) = (\pm\frac{7}{4}) \oplus (\pm\frac{1}{4})$.

The ghosts are a chiral vector and a real scalar superfield, plus the usual for the compactification vector multiplets:

$$\Lambda \otimes A_a = \Lambda_a, \qquad V \otimes C = K; \qquad \Lambda \otimes \varphi^I = \Lambda^I;$$

$$QU_a = (\Lambda_a + \bar{\Lambda}_a) + \partial_a K; \qquad QV^I = \Lambda^I + \bar{\Lambda}^I$$

The compensator (dilaton superfield) is a chiral scalar superfield (which identifies this supergravity as old minimal):

$$\Lambda \otimes \tilde{C} - \tilde{\Lambda} \otimes C = \phi,$$

$$Q\phi = \partial_a \Lambda^a + \bar{d}^2 d^2 K$$

This information is enough to write the low-energy action for the 4D N=1 heterotic string in the absence of Calabi-Yau matter. The procedure is to write a superconformally invariant action for the tensor (matter) multiplet and chiral scalar (compensator) multiplet, and then couple to conformal supergravity by general covariantization. It is useful to take the conformal weight of the compensator into account by considering also the cosmological term. This term is generally written as the chiral integral of $\phi^3$. However, for string purposes it is more convenient to make a field redefinition so that it appears as $\phi^2$; this means that $\phi$ now has conformal weight 3/2. The field strength $\tilde{G}$ of the tensor multiplet has weight 2, as fixed by the antisymmetric tensor gauge field $b_{ab}$ it contains. (This field strength also contains the contribution from the vector multiplets [8]; see the following subsection.) The fact that the dilaton appears to the same power (now 2) in both classical terms, together with superconformal invariance (which fixes the scale weight of both terms) now fixes the $\tilde{G}$ dependence:

$$S = \int d^4x\, d^4\theta\; \bar{\phi}\phi\tilde{G}^{-1/2} + \left(\int d^4x\, d^2\theta\; \tfrac{1}{2}\lambda\phi^2 + h.c.\right)$$

This is the case $n = -\tfrac{1}{2}$ of the "16−16" supergravity action considered in [7] (with the above field redefinition). This particular coupling to $\phi$ has the unique property that the supercurrent (defined by varying the action with respect to $U^a$), in the string gauge $\tilde{G} = 1$, has no spinor derivatives:

$$J_a = \tfrac{1}{2}\bar{\phi}i\overleftrightarrow{\nabla}_a\phi$$

(This is the case $n = 0$, $\tilde{n} = -\tfrac{1}{2}$ in the language of [17].) By using the appropriate covariantization (i.e., choice of vector connections), the current is also the field equation; the resulting simple equation $\nabla_a(ln\,\phi - ln\,\bar{\phi}) = 0$ appears naturally in the sigma-model approach [10]. We can now choose one of two scale gauges: The string gauge $\tilde{G} = 1$, or the standard gauge $\phi = 1$. The string gauge always gauges away the antisymmetric tensor gauge field, so that it is absorbed into the same prepotential as the metric tensor (as follows from the open $\otimes$ open argument, or the equivalent T-duality). Note that, if this action is expanded about the vacuum values of the fields, $\phi$ appears with the wrong sign (compensator), while $G$ appears with the right one (matter) [7].

While the real $d^4\theta$ term is generally written with a factor of $sdet(E_A{}^M)^{-1}$ to make it a density, this real factor cannot appear in the chiral $d^2\theta$ term, where $\phi^2$ acts as the corresponding density. In fact, all truly chiral superfields are such densities, with density weight corresponding directly to conformal weight. This simplifying feature of chiral integrals is one reason why their component evaluation is simpler than integrals over full superspace. A similar procedure can also be applied to nonsupersymmetric theories, with gravity written as conformal gravity plus compensating scalar: By writing all fields as densities, all factors of $\sqrt{-g}$ can be removed from the action. (A similar procedure was applied in the string example of section 2.2).

The compactification-dependent massless states can easily be added. Assuming that the compactification manifold has no isometries, all moduli are described by scalar fields. As is well known, N=1 scalar multiplets with general Yukawa couplings are described by chiral scalar superfields $\phi^I$. Furthermore, such fields must have conformal weight zero to allow 4D $\sigma$-model type actions (with invariances scalar $\rightarrow$ scalar + constant and nonlinear generalizations). This means the action must be of the form

$$S = \int d^4x\, d^4\theta\; \bar{\phi}\phi \tilde{G}^{-1/2} \mathcal{A}(\phi^I, \bar{\phi}^I) + \left[ \int d^4x\, d^2\theta\; \tfrac{1}{2}\phi^2 \mathcal{B}(\phi^I) + h.c. \right]$$

for some functions $A$ and $B$ (in terms of $\phi^I$ that are covariantly chiral with respect to the Yang-Mills fields as well as supergravity).

### 3.3. Super Chern-Simons forms

To cancel the usual one-loop anomalies, the tensor multiplet gauge transformation and field strength are modified. This modification can be described very simply in superspace. A related simpler problem, which we consider first, is to write a simple geometric form of the 4D N=1 super Yang-Mills action as a $d^4\theta$ integral. The first thing to consider is the Chern-Simons super-form, defined as the natural generalization of the bosonic expression to curved space [18]:

$$X_{MNP} = \tfrac{1}{2} A_{[M} \partial_N A_{P)} + \tfrac{1}{3} A_{[M} A_N A_{P)}$$

$$X_{ABC} = e_A{}^M e_B{}^N e_C{}^P X_{MNP} = \tfrac{1}{2} A_{[A} d_B A_{C)} - \frac{1}{4} A_{[A} T_{BC)}{}^D A_D + \tfrac{1}{3} A_{[A} A_B A_{C)}$$

where $d_A$ is the derivative covariantized with respect to just supergravity and not super Yang-Mills, and the usual grading sign factors are implicit.

This form is defined in arbitrary superspaces; an interesting set of special cases is those relevant for the classical superstring, namely N=1 superspace in D=3,4,6,10. We note that there the action (in a notation suitable for all dimensions)

$$S_{CS} = tr \int d^D x\, d^{2(D-2)}\theta\; \gamma^{a\alpha\beta} X_{a\alpha\beta}$$

which is the analog of the usual 3D bosonic Chern-Simons action with the Levi-Civita tensor replaced with a gamma matrix, is dimensionless in all these cases. In particular, in D=3 it gives a geometrical form of the usual super Chern-Simons action [18]. Furthermore, if we treat $A_a$ as independent from $A_\alpha$, then variation of $A_a$ imposes the usual conventional constraint $\gamma^{a\alpha\beta}F_{\alpha\beta} = 0$ in arbitrary dimensions, which makes it a sort of first-order action. (In D=3 this is the only constraint.) This is analogous to the harmonic superspace formulation of N=3 super Yang-Mills theory [5], where also one of the fields appears as auxiliary in a Chern-Simons action because of a nonvanishing (constant) torsion.

In D=4 this is the usual super Yang-Mills action, if we impose the representation-preserving constraints by hand or by Lagrange multipliers. This is easy to verify explicitly for the linearized action. For the fully nonlinear action, this is most easily seen by looking at the field equations: Including the conventional constraint coming from varying $A_a$ as described above, we have

$$\delta S_{CS} = tr \int d^D x d^{2(D-2)}\theta \, \gamma^{a\alpha\beta}(\delta A_{[a})F_{\alpha\beta)}$$

After solving the representation-preserving constraint, as

$$A_\alpha = \left(e^{-\gamma_5 V} d e^{\gamma_5 V}\right)_\alpha$$

we can write

$$\delta A_\alpha = -(\Delta V)(\gamma_5 \nabla)_\alpha$$

where

$$\Delta V \equiv e^{-V}\delta e^V$$

and $\nabla$ is the Yang-Mills covariantized derivative. The field equation is then

$$(\gamma_5\gamma)^{a\alpha\beta}\nabla_\alpha F_{\beta a} = 0$$

Using the identity

$$F_{\alpha a} = \gamma_{a\alpha\beta}W^\beta$$

which follows from the constraints, this can be written in the usual form

$$\nabla\gamma_5 W = 0$$

(The Bianchi identity on $W$ is $\nabla W = 0$.)

Another simple way to evaluate this expression is in the chiral representation, where $A_{\dot\alpha} = 0$, and only the $(A_a)^2$ term survives. In that representation $A_a \sim \bar{d}_{\dot\alpha} A_\alpha$ and $W_\alpha \sim \bar{d}^2 A_\alpha$, so one immediately obtains $\int d^2\theta \, W^2$.

As a first-order action, $A_a$ contains the Yang-Mills field strength (at order $\theta\bar\theta$) as an independent auxiliary field, so this action contains the usual first-order action for nonsupersymmetric Yang-Mills theory. (Similar remarks apply to the harmonic superspace formulation of N=3 super Yang-Mills theory.) The fact that such an expression exists for the N=1 super Yang-Mills action as a $d^4\theta$ integral in terms of potentials $A_A$ without explicit prepotentials $V$ is why this action gets renormalized at more than one loop, since in the background field method only such terms can occur in the effective action. Such an expression does not exist for N=2 super Yang-Mills theory, which is why it is finite at two loops and higher.

The generalization of the tensor-multiplet field strength is now easy to guess:

$$\tilde H_{ABC} = H_{ABC} + X_{ABC}$$

where $H_{ABC}$ is the usual super 3-form field strength for the tensor multiplet. In particular, we have

$$\tilde G = G + tr\,\gamma^{a\alpha\beta}X_{a\alpha\beta}$$

from which it follows that

$$(\bar d^2 + R)\tilde G \sim tr\,W^2$$

by calculations similar to those that showed the equivalence of the Chern-Simons form of the Yang-Mills action to the usual $\int d^4x d^2\theta\, W^2$ form. (Thus, in the string gauge $\tilde G = 1$, $R \sim tr\,W^2$.)

## 4. N=2

### 4.1. Smaller superspaces

As for N=1, all the important multiplets except conformal supergravity are described by superforms. Of the three propagating superforms, two can best be described in slightly different versions of harmonic superspace [19], the third in chiral superspace:

(1) The simplest case is vector multiplets. Conformal actions can be written with the chiral field strengths $W^I$ as integrals over chiral superspace

$$S_{VM} = \int d^4x\, d^4\theta\, f(W^I) + h.c.$$

with the only requirement that $f$ be homogeneous of degree 2 for invariance under R-symmetry, which implies conformal symmetry. (R-symmetry transforms $\theta^i \to e^{i\zeta}\theta^i$, $W^I \to e^{2i\zeta}W^I$.) For example, the superconformal action for a single vector multiplet (abelian or nonabelian) is

$$S \sim tr \int d^4x\, d^4\theta\, \tfrac{1}{2}W^2$$

(2) For tensor multiplets, we introduce as complex bosonic coordinates the SU(2) doublet $u^i$ parametrizing the space CP(1), and actions have not only SU(2) invariance but also the local complex scale (projective) invariance $u^i \to \lambda u^i$, which allows us to choose the gauge $u^i = (1, z)$. Integration over $u$ is over a contour as $\oint u_i du^i$, which becomes the usual $\oint dz$ in this gauge. A tensor multiplet has a finite number of fields, and the explicit $u$ dependence of its field strength is simply

$$L_{++}(u) = \tfrac{1}{2}u^i u^j L_{ij}$$

One then defines the SU(2)-invariant spinor derivative

$$d_{+\alpha} = u^i d_{i\alpha}, \qquad \bar{d}_{+\dot\alpha} = u^i \bar{d}_{i\dot\alpha}$$

with respect to which $L_{++}$ is defined to be "analytic":

$$d_+ L_{++} = \bar{d}_+ L_{++} = 0$$

which implies the usual tensor-mutliplet Bianchi identity

$$d_{(i\alpha} L_{jk)} = \bar{d}_{(i\dot\alpha} L_{jk)} = 0$$

This description is not only sufficient for the usual harmonic superspace manipulations, but the only one that manifests conformal invariance, which is particularly important for coupling to conformal supergravity. The natural form of superspace integration is then [20]

$$\int d^4\theta_\natural \equiv \oint u_i du^i \int \left(\frac{v_i d\theta^i}{v_j u^j}\right)^4$$

The $\theta$ integral is $v$-independent if the integrand is analytic. (For example, we can pick $v_i = \delta_{i+}$ to get a "twisted chiral" integral.) Superconformal actions for tensor multiplets $L^I_{++}$ are then

$$S_{TM} = \int d^4x\, d^4\theta_\natural\, f(L^I_{++})$$

where $f$ is homogeneous of degree 1 for projective invariance, which implies superconformal invariance.

(3) For scalar multiplets, we use also the complex conjugate coordinates $\bar{u}^i$, and constrain both by the condition $u^i \bar{u}_i = 1$. The local invariance is then only the U(1) phase transformation on $u$, so the space is SU(2)/U(1) (which is effectively the same as CP(1), but an invariance has been replaced by a constraint). Integration $\int du$ over $u$ and $\bar{u}$ is now defined to pick out the SU(2) singlet in terms of $u$-$\bar{u}$ dependence. Use of $\bar{u}$ allows us to define the generators of a second (broken) SU(2)

$$d_{++} = u^i \frac{\partial}{\partial \bar{u}^i}, \qquad d_{--} = \bar{u}^i \frac{\partial}{\partial u^i}, \qquad d_{+-} = u^i \frac{\partial}{\partial u^i} - \bar{u}^i \frac{\partial}{\partial \bar{u}^i}$$

and to write the other half of the spinor derivatives as

$$d_{-\alpha} = \bar{u}^i d_{i\alpha}, \qquad \bar{d}_{-\dot{\alpha}} = \bar{u}^i \bar{d}_{i\dot{\alpha}}$$

The concept of analytic superfields can then be extended:

$$d_+ L_{(n)} = \bar{d}_+ L_{(n)} = (d_{+-} - n) L_{(n)} = 0$$

However, such superfields can in general contain an infinite number of auxiliary fields. The superspace integration that includes $\int du$ is

$$\int d^4 \theta_\sharp \equiv \int du \int (\bar{u}_i d\theta^i)^4$$

We can also write the $u$ integration as

$$\int du\, f \equiv [1 - (d_{++})^{-1} d_{++}] f \quad \text{when} \quad d_{+-} f = 0, \quad 0 \quad \text{otherwise}$$

where $(d_{++})^{-1}$ is defined as 0 on states annihilated by $d_{--}$, and the inverse of $d_{++}$ otherwise. (This operator can be defined for general representations of SU(2), and is also useful in first-quantized BRST [21].) When applying this operator on analytic superfields, it is useful to remember that $L_{(n)}$ (for $n > 0$) contains only isospins $\geq n/2$ (with respect to the second SU(2)). The scalar multiplet has $(d_{+-})$ U(1) charge $n = 0$, and its superconformal actions are

$$S_{SM} = \int d^4x\, d^4\theta_\sharp\; f_{JK}(L^I)(d_{++} L^J)(d_{++} L^K)$$

for U(1) invariance, which implies superconformal invariance.

## 4.2. Versions of off-shell supergravity

These actions generalize directly to curved superspace by covariantizing the definition of harmonic and chiral superfields: Just as for chiral integrals in N=1 superspace, there is no factor involving the determinant of the vielbein.

As a consequence of simple dimensional analysis, all the classical terms are integrals over these smaller superspaces. Furthermore, the conformal weights of vector and tensor multiplets are determined by their gauge fields: $W$ has weight 1, while $L_{ij}$ has weight 2. More generally, superconformal weights of (twisted) chiral superfields are not arbitrary: The representation of superconformal symmetry in chiral superspace [22] implies that $(4-N)\times$(conformal weight) $= N\times$(U(1) weight), and that the chiral superfield have no external isospin or undotted spinor indices. This also follows from considering the superconformal transformation of $d_{i\alpha}$, using the fact that the chirality condition $\bar{d}_{i\dot\alpha} = 0$ must be preserved. From similar considerations of the analyticity constraint $u^i d_{i\alpha} = u^i \bar{d}_{i\dot\alpha} = 0$, we find that analytic superfields must have no external SU(2) or Lorentz indices, their (R-symmetry) U(1) charges must vanish, and their conformal weights are just their order in $u - \bar{u}$ (the $d_{+-}$ U(1) charge). This is consistent with what we know for the tensor multiplet field strength (and all functions of tensor multiplet field strengths), and implies the scalar multiplet field strength has vanishing conformal weight, as appropriate for a field that can be used to describe 4D nonlinear $\sigma$ models.

The table for N=2 superforms is:

| rank | reduced $A$ | reduced $dA$ | superhelicity | super(iso)spin |
|---|---|---|---|---|
| 0 | $L$ | $d_{++}L$ | 0 | 0 (1,2,...) |
| 1 | $L_{++}$ | $\int du\ (\bar{d}_-)^2 L_{++}$ | $\frac{1}{2}$ | 0 (0) |
| 2 | $\Phi$ | $(d_+)^2\Phi + (\bar{d}_+)^2\bar\Phi$ | 0 | 0 (0) |
| 3 | $L_{++}$ | $d_{++}L_{++}$ | — | 0 (1,2,...) |
| 4 | $L_{+++}$ | 0 | — | — |

$\Phi$ is chiral, while the $L$'s are analytic, with U(1) weight as indicated by the "+" signs.

While all these multiplets (except the trivial 4-form) describe superspin 0, the vector and tensor multiplet each describe a single superisopsin (0), while the others describe an infinite number (1,2,...). Thus the vector and tensor multiplets each can be written in terms of a single ordinary superfield, and are the only ones relevant for the N=(1,1) case, while the scalar multiplet will be important for N=(2,0). The table gives our previous expression for the field strength of the tensor multiplet with

$$L_{ij} = d_{ij}^2 \Phi + \bar{d}_{ij}^2 \bar\Phi$$

while the (off-shell dual) relationship for the vector multiplet is that we can write (in a particular non-derivative gauge)

$$A_{++} = (d_+)^4 \tfrac{1}{2} \bar{u}_i \bar{u}_j V^{ij}, \qquad W = \bar{d}^4 \tfrac{1}{2} d^2_{ij} V^{ij}$$

($(d_+)^4$ means $(\bar{d}_+)^2 (d_+)^2$.) The Yang-Mills field appears in the covariant derivative $\nabla_{++} = d_{++} + A_{++}$ [19].

The conformal supergravity multiplet (superspin 1, superisospin 0) can also be described by a single ordinary superfield. (Superspin $\tfrac{1}{2}$ is now the spin-$\tfrac{3}{2}$ mulitplet, excluded except for N>2 supersymmetry.) The local U(1) R-symmetry of the N=1 case is now generalized to U(2). Various sets of auxiliary fields for 4D N=2 supergravity have appeared in the literature. They all use the vector multiplet as a compensator for scale and U(1) transformations, but different versions of the scalar multiplet as a compensator for SU(2) transformations [12,19]: (1) the "nonlinear" tensor multiplet, (2) a partly on-shell version of the scalar multiplet, (3) the tensor multiplet, and (4) the harmonic scalar multiplet. The former two do not have a formulation in terms of unconstrained superfields (prepotentials), and so are not of general use, while the latter two are cases of the N=2 superforms just described. Since the scalars of the tensor multiplet form an isovector of SU(2), while those of the scalar multiplet form an isospinor, the former spontaneously breaks SU(2)→U(1), while the latter breaks SU(2) completely. They are therefore the analogs of the new minimal and old minimal cases of N=1 supergravity. The actions in these two cases are

$$S_{\text{``OM''}} = \int d^4x\, d^4\theta\, \tfrac{1}{2} W^2 + \int d^4x\, d^4\theta_\sharp\, \tfrac{1}{2}(d_{++} L)^2$$

$$S_{\text{``NM''}} = \int d^4x\, d^4\theta\, \tfrac{1}{2} W^2 + \int d^4x\, d^4\theta_\sharp\, L_{++} \ln L_{++} + \lambda \left( \int d^4x\, d^4\theta\, \Phi W + h.c. \right)$$

The cosmological term can also be written as $\int d^8\theta\, V^{ij} L_{ij}$, or as $\int d^4\theta_\sharp\, A_{++} L_{++}$.

## 4.3. N=(1,1)

For the N=2 type II string, we consider the direct product of 2 N=1 vector multiplets. The physical superfield is then just a real N=2 scalar superfield:

$$V(\theta^+, \bar{\theta}_+) \otimes V(\theta^-, \bar{\theta}_-) = U(\theta^+, \bar{\theta}_+, \theta^-, \bar{\theta}_-)$$

(The ± indices are now for the physical SU(2); we have not yet introduced the harmonic coordinates $u^i$.) The superspin analysis is now $\tfrac{1}{2} \otimes \tfrac{1}{2} = 1 \oplus 0$ (in terms of N=1 superspins in the left half of the equation, N=2 in the right), describing again conformal supergravity plus a tensor multiplet. The corresponding superhelicity analysis is $(\pm\frac{1}{2}) + (\pm\frac{1}{2}) = (\pm 1) \oplus 0^2$, where the two superhelicities 0 describe a single tensor multiplet (since it is a complex representation).

The use of a general, real, scalar, isoscalar superfield to describe N=2 supergravity [11] follows from harmonic superspace [19] if the analyticity condition on the covariant derivative is solved in a globally supersymmetric way:

$$[d_{+\alpha}, \nabla_{++}] = 0, \qquad \nabla_{++} = d_{++} + \tfrac{1}{2} H^{\alpha\beta}_{++}\partial_{\alpha\beta} + H^{\alpha}_{+++} d_{-\alpha} + H_{++++} d_{--}$$

$$\Rightarrow \qquad \nabla_{++} = d_{++} + \tfrac{1}{24} C^{\alpha\beta\gamma\delta}\{d_{+\alpha}, [d_{+\beta}, \{d_{+\gamma}, [d_{+\delta}, U d_{--}]\}]\}$$
$$= d_{++} + \tfrac{1}{2}[(d_+)^{2\alpha\beta} U]\partial_{\alpha\beta} + [(d_+)^{3\alpha} U]d_{-\alpha} + [(d_+)^4 U]d_{--}$$

(For convenience, we have used 6D spinor notation, where $\alpha$ is an SU*(4) index, which includes both $\alpha$ and $\dot\alpha$, and $\partial_{\alpha\beta}$ is the vector derivative. The other possible superfields in $\nabla_{++}$ can be completely gauged away by nonderivative transformations.)

The ghosts, being products of real superfields with chiral ones, are chiral in only half of the $\theta$ coordinates:

$$\Lambda(\theta^+) \otimes V(\theta^-, \bar\theta_-) = C_+(\theta^+; \theta^-, \bar\theta_-), \qquad V(\theta^+, \bar\theta_+) \otimes \Lambda(\theta^-) = C_-(\theta^+, \bar\theta_+; \theta^-)$$

The resulting gauge transformation is:

$$QU = C_+ + \bar C_+ + C_- + \bar C_-$$

The compensators are chiral and twisted chiral superfields (and their complex conjugates):

$$\Lambda(\theta^+) \otimes \tilde\Lambda(\theta^-) - \tilde\Lambda(\theta^+) \otimes \Lambda(\theta^-) = W(\theta^+, \theta^-)$$
$$\bar\Lambda(\bar\theta_+) \otimes \tilde{\bar\Lambda}(\bar\theta_-) - \tilde{\bar\Lambda}(\bar\theta_+) \otimes \bar\Lambda(\bar\theta_-) = \overline{W}(\bar\theta_+, \bar\theta_-)$$
$$\Lambda(\theta^+) \otimes \tilde{\bar\Lambda}(\bar\theta_-) - \tilde\Lambda(\theta^+) \otimes \bar\Lambda(\bar\theta_-) = L_{++}(\theta^+, \bar\theta_-)$$
$$\bar\Lambda(\bar\theta_+) \otimes \tilde\Lambda(\theta^-) - \tilde{\bar\Lambda}(\bar\theta_+) \otimes \Lambda(\theta^-) = \bar L^{++}(\bar\theta_+, \theta^-)$$

While the chirality condition on $W$ is covariant with respect to the SU(2) symmetry that mixes the 2 $\theta$'s, the twisted chirality doesn't seem to be, until we realize that $L_{++}$ and $\bar L^{++} = L_{--}$ form 2 components of an isotriplet:

$$d_{(i\alpha} L_{jk)} = \bar d_{(i\dot\alpha} L_{jk)} = 0 \quad \Rightarrow \quad d_{+\alpha} L_{++} = \bar d^-_{\dot\alpha} L_{++} = 0$$

The fact that $L_{+-}$ doesn't appear explicitly in the direct product construction is a reflection of the "mirror" symmetry that implies the invariance $\delta L_{+-} = constant$ in the string gauge [13] (see below). Of the cases we consider, this set of compensators has the only nontrivial superhelicity calculation, since in all other cases at least one of the Yang-Mills ghost is for N=0 or 2, both of which have superhelicity 0. (For superspin, all of N=0,1,2 have superspin 0.) The analysis in this case is $(\pm\frac{1}{4})+(\pm\frac{1}{4}) = (\pm\frac{1}{2}) \oplus 0^2$, the vector and tensor multiplets. Their gauge transformations are

$$QW = (\bar{d}^2 d^2)_- C_+ + (\bar{d}^2 d^2)_+ C_-, \qquad QL_{++} = (d^2 \bar{d}^2)_- C_+ + (\bar{d}^2 d^2)_+ \bar{C}_-$$

where the + derivatives here involve only $d_+$ and its complex conjugate $\bar{d}^+$, and similarly for the − derivatives.

This theory therefore consists of conformal supergravity coupled to a physical tensor multiplet $L'_{ij}$ and tensor and vector compensator multiplets $L_{ij}$ and $W$. In the string gauge we gauge away the physical tensor multiplet $L'_{ij}$ as $L'_{ij} = \delta_{ij}$. Gauging away the 2 components not proportional to $\delta_{ij}$ is accomplished by using the local SU(2) symmetry of conformal N=2 supergravity: We rotate the isovector to point in a fixed direction. The remaining component is gauged to 1 by Weyl invariance. In this gauge we have two U(1)'s remaining: one the original one of SU(2)⊗U(1), and the second from breaking SU(2)→U(1). In this gauge both physical multiplets, (conformal) supergravity and the tensor multiplet, are contained within the prepotential $U$. The gauge more commonly used in supergravity is to gauge away as much of the compensators as possible, by the conditions $W = 1$ (scale and U(1)) and $L_{ij} \sim \delta_{ij}$ (again breaking SU(2)→U(1)), leaving $L_{+-}$ unfixed.

As in the N=1 heterotic case, this information on the multiplet structure, together with the fact that in string theory compensators appear only quadratically in the low-energy action, is sufficient to determine this action in the absence of compactification matter:

$$S = \int d^4x\, d^4\theta_\natural\, \frac{\frac{1}{2}L_{++}^2}{L'_{++}} + \int d^4x\, d^4\theta\, \tfrac{1}{2}W^2 + \lambda\left(\int d^4x\, d^4\theta\, \Phi W + h.c.\right)$$

where $L_{++}$ and $L'_{++}$ are now again the harmonic superfields. Note that the contour integral is particularly simple in the basis $u^i = (1, z)$, where the string gauge is simply $L'_{++} = z$. We then just integrate around the pole at $z = 0$. This also means the action has an invariance [13]

$$\delta L_{++} \sim L'_{++}$$

(i.e., $\delta L_{++} \sim z$ in the string gauge) if the cosmological term is dropped. This symmetry is a consequence of the mirror symmetry that relates type IIA and IIB strings, by switching tensor and vector multiplets. Since each vector multiplet has only two scalars, this allows only two of the three scalars of each tensor multiplet to appear in the action without derivatives. In the string gauge this is the standard invariance $\delta L_{ij} = constant \times \delta_{ij}$.

Compactification matter can be added in a way similar to the N=1 case. Again the compactification-dependent massless states are described by supermultiplets containing scalars. For these to have general self-interactions, they must couple to the compensators. As described in the previous subsection, the self-interaction terms of vector and tensor multiplets consist of only a chiral term containing just vector multiplets, and a "harmonic analytic" term containing just tensor multiplets. These compactification multiplets must therefore themselves also be vector and tensor multiplets. Unlike N=1, this matter must be expressed as ratios of matter field strengths to compensator field strengths, to obtain dimensionless scalars:

$$S = \int d^4x\, d^4\theta_\natural\, \frac{\tfrac{1}{2}L_{++}^2}{L'_{++}} \mathcal{A}\left(\frac{L^I_{++}}{L_{++}}\right) + \left[\int d^4x\, d^4\theta\, \tfrac{1}{2}W^2 \mathcal{B}\left(\frac{W^I}{W}\right) + h.c.\right]$$

where we now have dropped the cosomological term to maintain the symmetry [13]

$$\delta L_{++} = kL'_{++}, \qquad \delta L^I_{++} = k^I L'_{++}$$

related to mirror symmetry, for arbitrary constants $(k, k^I)$. (For contributions to the string effective action from loops and high energy, see also [13].)

## 4.4. N=(2,0)

It should be possible to extend our arguments to closed strings that can be represented as direct products of open N=2 strings with other open strings. We then need to represent the vector multiplet and its ghost multiplet in N=2 superspace. The simplest way is to use harmonic superspace, since representing these ghosts as an analytic harmonic superfield avoids ghosts-for-ghosts. The linearized form of the BRST transformations for this vector multiplet (as follows, e.g., from the superform analysis above) is

$$QA_{++} = d_{++}L, \qquad QL = 0, \qquad Q\tilde{L} = (d_+)^4(d_{++})^{-2}(d_{--})^2(d_{++}A_{++})$$

BRST invariance of the gauge-fixed vector-multiplet action

$$\int d^4x\, d^4\theta_\natural\, [\tfrac{1}{2}A_{++}\Box A_{++} + \tilde{L}(d_{++})^2 L]$$

follows from applying the identities

$$(d_{++})^2(d_{++})^{-2}(d_{--})^2 d_{++} L_{++} = (d_{--})^2 d_{++} L_{++}, \qquad (d_+)^4 (d_{--})^2 L_{++} = \Box L_{++}$$

for an arbitrary analytic superfield $L_{++}$.

The first case is the heterotic N=2 string. The physical superfield is the direct product of a real analytic harmonic superfield with a vector:

$$A_{++} \otimes A^a = U^a_{++}$$

(As for the N=1 heterotic string, the vector multiplets arising from the 26→10 compactification are rather trivial as far as the open $\otimes$ open analysis is concerned, so we will not repeat it here.) As for the N=2 vector multiplet, this multiplet has only a finite number of fields: Just as we can express $A_{++}$ in terms of the ordinary superfield $V^{ij}$, we can also write $U^a_{++}$ in terms of an ordinary superfield $U^{ija}$. The superspin analysis is then $0 \otimes 1 = 1$, which implies this superfield is the irreducible conformal supergravity multiplet off shell. This would imply a new formulation of conformal supergravity, since previously the physical antisymmetric tensor gauge field was not contained in this muliplet. On the other hand, the superhelicity analysis is $(\pm\frac{1}{2}) + (\pm 1) = (\pm\frac{3}{2}) \oplus (\pm\frac{1}{2})$, describing supergravity plus a vector multiplet on shell. (A similar multiplet, with the same content on shell, but with 32+32 components off shell instead of 24+24, has been considered by Müller [23] in a component analysis derived from torsion constraints. A multiplet similar to Müller's, obtained by coupling the usual conformal supergravity to a vector multiplet with an off-shell central charge [24], was proposed in [25].) This multiplet is thus the opposite of the usual supergravity multiplets, which are reducible off shell (conformal supergravity + compensators), but irreducible on shell (superhelicity $2-\frac{1}{4}$N).

We also have the ghosts

$$L \otimes A^a = L^a, \qquad A_{++} \otimes C = L_{++}$$

and gauge transformation

$$QU^a_{++} = d_{++} L^a + \partial^a L_{++}$$

For the compensator we find

$$L \otimes \tilde{C} - \tilde{L} \otimes C = L$$

$$QL = \partial_a L^a + (d_+)^4(d_{++})^{-2}(d_{--})^2 d_{++} L_{++}$$

The compensator is then the same representation as the super Yang-Mills ghosts, a scalar multiplet with an infinite set of auxiliary fields.

By the same methods as for previous cases, the low-energy action (in the absence of compactification matter) is

$$S = \int d^4x d^4\theta_\sharp \tfrac{1}{2}(d_{++}L)^2$$

Although such an action has been proposed previously, we use a new formulation of conformal supergravity. It is then natural to assume that the vector multiplets from the 26→10 reduction can be included in a manner similar to that used for N=1: modification of the Bianchi identity of a conformal multiplet. In the N=1 case, we had $(\bar{d}^2 + R)\tilde{G} \sim tr\, W^2$. If we look at the N=1 superform table (section 3.1), it is clear that only the tensor multiplet has a chiral scalar Bianchi identity, necessary since the vector multiplet field strength is chiral. That field strength is chiral also for N=2, but in that case the only chiral scalar Bianchi identity is for the scalar multiplet (from the table in section 4.2). The result is then

$$S = \int d^4x d^4\theta_\sharp \tfrac{1}{2}\tilde{G}_{++}{}^2$$

where

$$\tilde{G}_{++} = d_{++}L + \Omega_{++}, \qquad \int du(\bar{d}_-)^2 \Omega_{++} \sim tr\, W^2$$

(Unlike the N=1 case, the Chern-Simons term $\Omega$ does not allow the existence of higher-loop renormalization of Yang-Mills theory, since the Yang-Mills action is represented in terms of it as an integral over only 6 $\theta$'s and not the full 8.)

In particular, we can compare the component fields of this version of conformal supergravity to the usual one, by taking the direct product in terms of the component fields, $(A_a, \chi_{i\alpha}, D_{ij}) \otimes A_a$: Besides the obvious conformal graviton and gravitini, we have

| old | new |
|---|---|
| $W_{\alpha\beta}$ (1) | $A_a, V_a$ (0) |
| $G_{(ij)a}$ (1) | $G_{(ij)a}$ (1) |
| $G_a$ (1) | $b_{ab}$ (0) |
| $\lambda_{i\alpha}$ ($\tfrac{3}{2}$) | $\chi_{i\alpha}$ ($\tfrac{1}{2}$) |
| $D$ (2) | $\varphi$ (0) |

where the dimension (conformal weight) is indicated in parentheses. The auxiliary antisymmetric tensor has been replaced with two gauge vectors, one of which can be identified on shell with the physical vector of supergravity, the other with that of the vector multiplet. The Lagrange multipliers of dimension $\frac{3}{2}$ and 2 have been replaced with fields of physical dimension: These, together with the tensor gauge field, which replaces the U(1) gauge field, describe the remaining physical degrees of freedom of the vector multiplet on shell. By comparison, the scalar multiplet compensator lacks a compensator for U(1), but now it only needs scalars that compensate SU(2) and scale. Since there is a single compensator multiplet, this formalism also lacks the doubling of dilatons and dilatini for which the Lagrange multipliers were necessary.

## 5. N>2

### 5.1. N=(2,1)

Another interesting type of string can result only from asymmetric compactification of the type II string, so that different numbers of supersymmetries survive in the left- and right-handed sectors (such as with, e.g., asymmetric orbifolds [26]). The physical superfield is a direct product of an analytic N=2 superfield with a real N=1 superfield:

$$A_{++} \otimes V = U_{++}$$

This real scalar superfield is thus analytic in the first two $\theta$'s and general in the third. This is the same type of superfield used in the harmonic superspace formulation of N=3 super Yang-Mills [5]. However, because of its U(1) weight and gauge transformations, this superfield contains only a finite number of fields, like the case of N=(2,0) conformal supergravity, but unlike the N=3 Yang-Mills case. We can then also express this superfield in terms of an ordinary N=3 superfield $U^{ij}$, where $i,j$ are SU(2) (not SU(3)) indices. The superspin analysis is now $0 \otimes \frac{1}{2} = \frac{1}{2}$. This is again the conformal supergravity multiplet. The superhelicity is $(\pm\frac{1}{2}) + (\pm\frac{3}{4}) = (\pm\frac{5}{4}) \oplus (\pm\frac{1}{4})$, which is again supergravity plus a vector multiplet. (The supergravity multiplet has helicities running from 2 to $\frac{1}{2}$, and the complex conjugate states, while the vector multiplet runs from 1 to $-\frac{1}{2}$, and complex conjugates.) The situation is thus very similar to N=(2,0).

The ghosts are now

$$L \otimes V = L', \qquad A_{++} \otimes \phi = \Upsilon_{++}$$

where $L'$ has similar properties to $U_{++}$ (except for U(1) weight), and $\Upsilon_{++}$ is analytic in the first two $\theta$'s but chiral in the third. While such harmonic superfields have been considered as representations of (conformal) N=3 supersymmetry, they have not been used to describe physical multiplets. The gauge transformation is then

$$QU_{++} = d_{++}L' + (\Upsilon_{++} + \bar{\Upsilon}_{++})$$

(We use the harmonic complex conjugate, which preserves $u$ and $\bar{u}$ instead of switching them.)

The compensator is

$$L \otimes \tilde{\phi} - \tilde{L} \otimes \phi = \Upsilon$$

and its gauge transformation is

$$Q\Upsilon = (\bar{d}^2 d^2)_3 L' + (d_+)^4 (d_{++})^{-2} (d_{--})^2 d_{++} \Upsilon_{++}$$

This $\Upsilon$ is similar to $\Upsilon_{++}$ except for U(1) weight. Again, such superfields have not previously been applied to physical multiplets; however, the compensators for N=3 supergravity are expected to be vector multiplets: First, the vector multiplet is the only one with spins no higher than 1, and the N=3 conformal supergravity multiplet already has the appropriate set of spin 2 and $\frac{3}{2}$ fields for N=3 supergravity. Second, the vector multiplet's scalars form a 3 of SU(3), so a $3 \oplus \bar{3}$ of vector multplets give $3 \otimes \bar{3} = 1 \oplus 8$ (and $3 \otimes 3 = \bar{3} \oplus 6$), which are the dilaton (1) and compensators for the SU(3) gauge fields (8). We therefore expect $\Upsilon$ to describe a new off-shell representation of vector multiplets.

## 5.2. N=(2,2)

The last example we consider has N=4 supersymmetry from the direct product of two N=2 supersymmetries. The resulting physical superfield is harmonic analytic in both pairs of $\theta$'s:

$$A_{++} \otimes A_{+'+'} = U_{+++'+'}$$

where we use $\pm$ as the indices of the first broken harmonic SU(2) and $\pm'$ for the second. As in all the other cases, the physical superfield contains a finite number of physical and auxiliary fields (because so do the N=0,1,2 vector multiplets). The superspin analysis is $0 \otimes 0 = 0$, again conformal supergravity. The superhelicity decomposition is $(\pm\frac{1}{2}) + (\pm\frac{1}{2}) = (\pm 1) \oplus 0^2$, describing supergravity plus two vector multiplets. (Supergravity has helicities $2, \frac{3}{2}, 1, \frac{1}{2}, 0$ and complex conjugates, while the vector multiplet has $1, \frac{1}{2}, 0, -\frac{1}{2}, -1$ and is a real representation.)

The ghosts are

$$L \otimes A_{+'+'} = L_{+'+'}, \qquad A_{++} \otimes L' = L_{++}$$

Both are similar to the physical superfield, but have different weights under the two U(1)'s, resulting in their containing an infinite number of auxiliary fields. The gauge transformation of the physical superfield is

$$QU_{+++'+'} = d_{+'+'}L_{++} + d_{++}L_{+'+'}$$

The compensator is

$$L \otimes \tilde{L}' - \tilde{L} \otimes L' = L''$$

and its gauge transformation is

$$QL'' = (d_{+'})^4(d_{+'+'})^{-2}(d_{-'-'})^2 d_{+'+'}L_{+'+'} + (d_+)^4(d_{++})^{-2}(d_{--})^2 d_{++}L_{++}$$

With regard to earlier difficulties in finding an off-shell N=4 formulation of supergravity, one should take into account that the N=4 strings come with particular choices of matter multiplets: For example, straight dimensional reduction of N=1 supergravity from D=10 yields N=4 supergravity plus six vector multiplets. On the other hand, the compensators for 4D N=4 supergravity are also six vector multiplets: The vector multiplet's scalars form a 6 of SU(4), and 6⊗6 = 1⊕15⊕20 gives the dilaton singlet, as well as the Stueckelberg field for the auxiliary vector of conformal supergravity that gauges SU(4). (As for N=2, the six compensating vectors are the physical vectors of N=4 supergravity.) This doubling of N=4 vector multiplets, with opposite-sign kinetic terms, is exactly what is needed to avoid the no-go theorem for the off-shell formulation of the N=4 vector multiplet [27].

## ACKNOWLEDGMENTS

I thank Jan de Boer and Gordon Chalmers for discussions on asymmetric compactification, Jim Gates for references, and Nathan Berkovits for discussions and especially for encouraging me to write this paper. This work was supported in part by the National Science Foundation Grant No. PHY 9309888.